\pdfoutput=1
\documentclass[10pt,conference]{IEEEtran}
\usepackage{cite}
\usepackage{amsmath,amssymb,amsthm,bm,booktabs}
\usepackage{graphicx}
\usepackage[caption=false,font=footnotesize]{subfig}
\usepackage[T1]{fontenc}
\ifdefined\pdfminorversion\pdfminorversion=7\fi
\newcommand{\E}{\mathbb E}
\newcommand{\CN}{\mathcal{CN}}
\newcommand{\MSE}{\operatorname{MSE}}
\newcommand{\CRB}{\operatorname{CRB}}
\newcommand{\Cov}{\operatorname{Cov}}
\newcommand{\tr}{\operatorname{tr}}

\newcommand{\Ree}{\operatorname{Re}}
\newcommand{\Imm}{\operatorname{Im}}
\newcommand{\MF}{\mathrm{MF}}
\newcommand{\RF}{\mathrm{RF}}
\newtheorem{theorem}{Theorem}
\newtheorem{corollary}{Corollary}

\begin{document}
	
	\title{Localization MSE Analysis for Cooperative OFDM-ISAC}
	
	\author{
		\IEEEauthorblockN{
			Kaitao Meng\IEEEauthorrefmark{1},
			Kawon Han\IEEEauthorrefmark{2},
			and Christos Masouros\IEEEauthorrefmark{3}
		}
		
		\IEEEauthorblockA{
			\IEEEauthorrefmark{1}The University of Manchester, Manchester, U.K.
		}
		
		\IEEEauthorblockA{
			\IEEEauthorrefmark{2}Ulsan National Institute of Science and Technology, Ulsan, South Korea
		}
		
		\IEEEauthorblockA{
			\IEEEauthorrefmark{3}University College London, London, U.K.
		}
	
	\IEEEauthorblockA{
		Emails: 
		kaitao.meng@manchester.ac.uk; kawon.han@unist.ac.kr; c.masouros@ucl.ac.uk 
	}
	
	}
	
	\maketitle
	
	\begin{abstract}
		This paper develops a receiver-specific mean-squared-error (MSE) framework for cooperative integrated sensing and communication (ISAC) localization using random orthogonal frequency-division multiplexing (OFDM) payloads. It captures local peak perturbations and incorrect ambiguity-peak acquisition for two classical receiver processing schemes: matched filtering (MF) and reciprocal filtering (RF). For scatterers well separated in delay, we quantify how frequency spread and interference induced by symbol-energy fluctuations affect local delay errors under MF. RF removes symbol-dependent weighting but amplifies noise for low-energy symbols. Payload-conditioned joint Gaussian delay profiles yield peak-selection probabilities, which combine with nonlinear multistatic fusion errors to approximate position MSE while accounting for network geometry. Numerical results illustrate receiver-dependent estimation errors and the benefits of frequency-dependent constellation allocation.
	\end{abstract}
%	\begin{IEEEkeywords}
%		ISAC, OFDM, MSE, matched filtering, reciprocal filtering, cooperative localization.
%	\end{IEEEkeywords}
	
	\vspace{-1mm}
	\section{Introduction}
	\vspace{-1mm}
	Network-level integrated sensing and communication (ISAC) coordinates geographically distributed base stations (BSs) to extend coverage and support cooperative localization \cite{Li2023Towards,Meng2024CooperativeTWC}. Reusing orthogonal frequency-division multiplexing (OFDM) communication payloads for sensing can facilitate ISAC deployment on existing cellular infrastructure. In this setting, ranging accuracy depends jointly on the random OFDM payloads, transmit powers, and receiver processing \cite{Zhang2021OverviewSignal}. An accurate yet computationally tractable performance metric is therefore needed to guide waveform design and resource allocation. The Cram\'er--Rao bound (CRB) benchmarks
	local information and geometry \cite{godrich2010target,Liu2022Limits},
	but does not capture incorrect ambiguity-peak acquisition.
	The Ziv--Zakai bound (ZZB) captures ambiguity-induced threshold
	behavior through binary hypothesis testing
	\cite{ZivZakai1969,Bell1997}. However, the probability integrals and averaging over random payloads required in this setting complicate its use as an optimization criterion.

\begin{figure}[t]
		\centering
		\includegraphics[width=7.5cm]{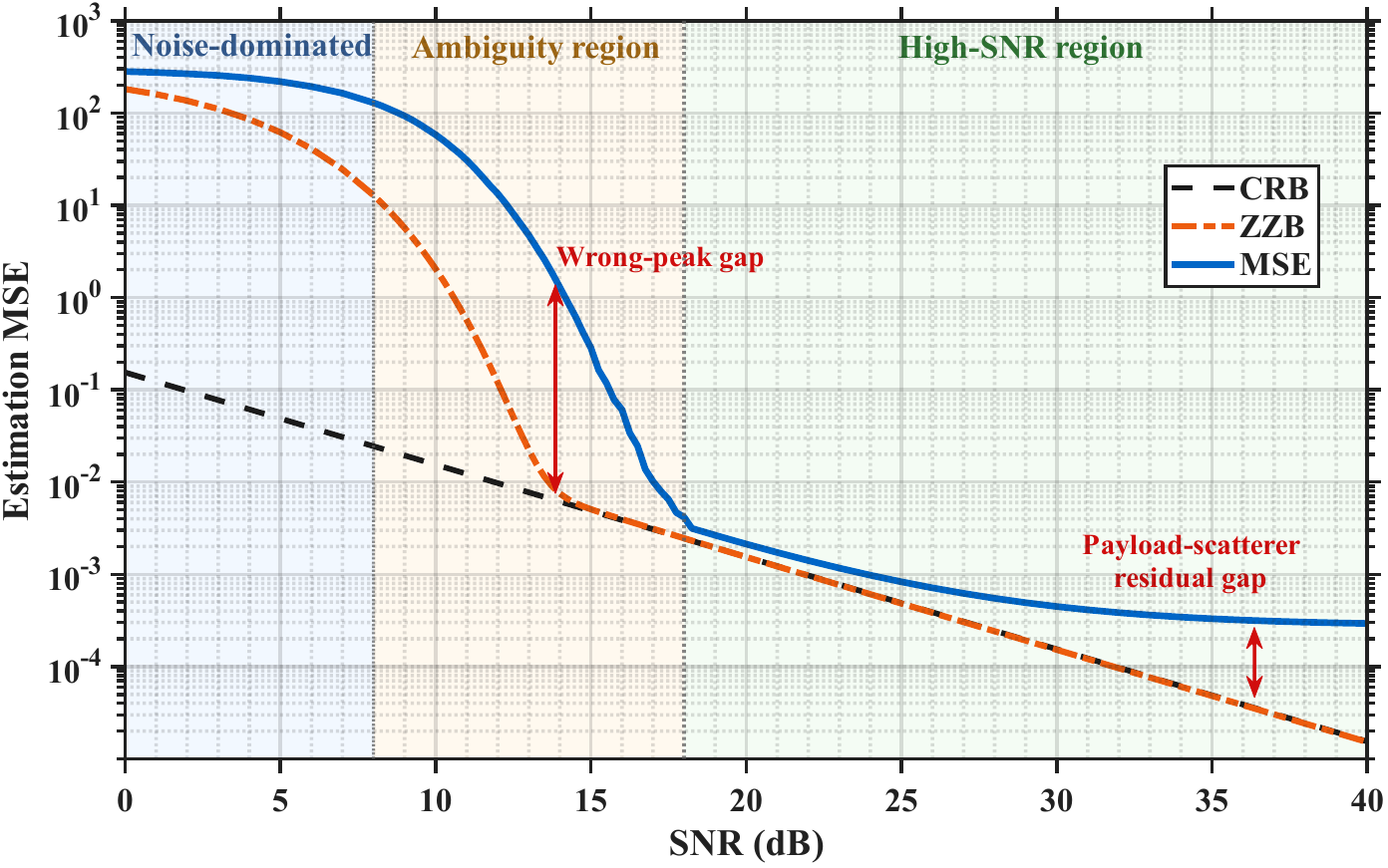}
		\vspace{-3mm}
		\caption{Conceptual MSE regimes in delay estimation.}
		\label{fig:mse_regimes}
	\end{figure}

	Fig.~\ref{fig:mse_regimes} schematically illustrates why the CRB and ZZB may fail to accurately characterize the mean-squared error (MSE) achieved by a practical OFDM sensing receiver. At low signal-to-noise ratio (SNR), noise dominates and the finite delay-search interval constrains the estimation error. Near the acquisition threshold, incorrect ambiguity-peak selection can produce large errors and a substantial gap from the local CRB. Although the ZZB captures this threshold behavior, it does not directly characterize a prescribed receiver's peak-selection statistics and therefore need not closely track its achieved MSE. At high SNR, once acquisition becomes reliable, local peak perturbations govern accuracy. Two commonly used receiver processing schemes are matched filtering (MF) and reciprocal filtering (RF), whose distinct error mechanisms can also leave a gap between the achieved MSE and the local CRB. With amplitude-varying payloads, MF produces payload-dependent sidelobes through symbol-energy weighting. Their coupling to undesired scatterers can perturb the desired peak and leave a residual error floor even as noise vanishes. RF removes this signal weighting but amplifies noise when symbol energies or transmit powers are small. These receiver-dependent effects influence both local accuracy and acquisition reliability, motivating a regional MSE analysis that accounts for local errors and the probabilities and displacement costs of incorrect peak selection.
	
	Receiver-dependent local MSE and constellation shaping are studied in \cite{Han2026MSE}, and finite-alphabet constellation selection and power control in \cite{Meng2026Constellation}, primarily for single-node sensing. Local accuracy alone cannot capture the displacement costs of rare incorrect peaks. Interval estimation addresses threshold behavior by partitioning the parameter domain \cite{Mallat2014TOA}. Particularly, cooperative sensing adds two challenges: shared payloads can couple acquisition outcomes across links, and incorrect delays can invalidate local fusion linearization. The resulting position errors depend on the affected links, network geometry, and nonlinear fusion.
	
	We develop a receiver-specific MSE framework that combines joint peak-selection probabilities from payload-conditioned Gaussian delay profiles with nonlinear fusion to approximate position MSE. The analysis retains correlations among candidate peaks and dependence between links sharing a payload. For fixed constellations, local MSE criteria yield closed-form RF power allocations under subcarrier bounds and a convex MF subproblem with a fixed frequency centroid.
	
	\section{System Model}
	\subsection{Geometry and Random OFDM Payloads}
	Consider base stations (BSs) at $\bm b_i\in\mathbb R^D$, $D\in\{2,3\}$,
	and a designated target $k$ at $\bm r_k$, where $i$ and $k$ are indices of BSs and targets. Let $\mathcal E$ be the prescribed
	set of directed sensing links and $E=|\mathcal E|$. For
	$e=(i,j)$, write $i_e=i$ and $j_e=j$ for notational convenience. The propagation length and delay via target \(k\) are
	\begin{equation}
		d_{e,k}=\|\bm r_k-\bm b_i\|+\|\bm r_k-\bm b_j\|,
		\qquad \tau_{e,k}=d_{e,k}/c .
		\label{eq:geometry}
	\end{equation}
	Here $c$ is the propagation speed. BS $i$ uses a prescribed active set
	$\mathcal N_i$ of $N_i=|\mathcal N_i|$ subcarriers at frequencies $f_{i,n}$,
	with $x_{i,n}=\sqrt{p_{i,n}}u_{i,n}$, $p_{i,n}>0$.
	Each active set contains at least two distinct frequencies.
	The symbols are independent across subcarriers and transmitting BSs, with
	$\E[u_{i,n}]=0$ and $\E[|u_{i,n}|^2]=1$. We define the symbol power variance and the inverse second moment, respectively, as
	\begin{equation}
		\kappa_{i,n}=\E[|u_{i,n}|^4]-1,\qquad
		\eta_{i,n}=\E[|u_{i,n}|^{-2}].
		\label{eq:moments}
	\end{equation}
	The  observation on link $e$ is denoted by
	\begin{equation}
		y_{e,n}=x_{i,n} \sum\nolimits_{\ell=1}^K\alpha_{e,\ell}
		e^{-j2\pi f_{i,n}\tau_{e,\ell}} +z_{e,n},
		\label{eq:observation}
	\end{equation}
	where $z_{e,n}\sim\CN(0,\sigma_e^2)$ are independent across links and
	subcarriers, and independent of the payloads and scene, with
	$\E[|z_{e,n}|^2]=\sigma_e^2$. Here $\alpha_{e,\ell}$ is a complex scattering coefficient, and
	the $K$ scatterers include targets and discrete clutter. We assume
	calibrated inter-BS timing, known payloads at the sensing receivers, and
	separable BS transmissions, as in OFDM cellular sensing \cite{Shi2022DeviceFree}.
	
	\subsection{Delay Search and Nonlinear Fusion}
The delay profiles for MF and RF processing are respectively given by
	\begin{align}
		s_e^{\MF}(\tau)&=\sum_{n\in\mathcal N_i}y_{e,n}x_{i,n}^*
		e^{j2\pi f_{i,n}\tau},\notag\\
		s_e^{\RF}(\tau)&=\sum_{n\in\mathcal N_i}\frac{y_{e,n}}{x_{i,n}}
		e^{j2\pi f_{i,n}\tau}.
		\label{eq:profiles}
	\end{align}
	A target-specific search interval $\mathcal T_{e,k}$ is prescribed
	independently of the current observation and lies within an unambiguous
	delay interval. The coarse estimate $\widetilde\tau_{e,k}$ maximizes
	$|s_e(\tau)|^2$ over a fixed oversampled grid
	$\mathcal G_e\cap\mathcal T_{e,k}$. With prescribed refinement half-width
	$\Delta_{\rm ref}>0$, it is refined continuously as
	\begin{equation}
		\widehat\tau_{e,k}=\arg\max_{\substack{\tau\in\mathcal T_{e,k}\\
				|\tau-\widetilde\tau_{e,k}|\leq\Delta_{\rm ref}}}|s_e(\tau)|^2.
		\label{eq:estimator}
	\end{equation}
	The delays are fused over a prescribed position domain $\Omega\subseteq\mathbb R^D$
	containing $\bm r_k$, using a fixed $\bm W\succeq\bm0$:
	\begin{equation}
		\widehat{\bm r}_k=\arg\min_{\bm r\in\Omega}
		\|c\widehat{\bm\tau}_k-\bm d(\bm r)\|_{\bm W}^2 ,
		\label{eq:fusion}
	\end{equation}
	where $\|\bm v\|_{\bm W}^2=\bm v^T\bm W\bm v$. Here $[\bm d(\bm r)]_e=\|\bm r-\bm b_i\|+\|\bm r-\bm b_j\|$.
	Here $\Omega$ is the known region of interest (ROI), or $\mathbb R^D$ if unconstrained.
	The prescribed $\mathcal T_{e,k}$ uses prior association/tracking information
	and fixes the delay ambiguity order, and a position ROI alone need not resolve it.
	The true-delay-centered region introduced next is used only to classify
	estimation outcomes. 
	
	\section{Single-Link Region-Wise Delay MSE}
	\label{SingleLink}
	
	We suppress link indices, write $N$ for the active-tone count, and use
	$q\in\{\MF,\RF\}$ for the receiver. The payload vector is $\bm u=[u_n]_n$.
	\subsection{Payload-Conditioned Information Reference}
	For an isolated target with unknown deterministic complex amplitude, set
	$q_n=p_n|u_n|^2$, $P_x=\sum_nq_n$, and
	$\bar f_x=P_x^{-1}\sum_nq_nf_n$. Eliminating the amplitude nuisance
	parameters by a Schur complement gives the delay information
	\begin{align}
		\mathcal J_{\tau_k|\alpha_k}(\bm u)
		=\frac{8\pi^2|\alpha_k|^2}{\sigma^2}J_x,
		\label{eq:conditional_information}
	\end{align}
	where $
	J_x=\sum_nq_n(f_n-\bar f_x)^2.$ The subtraction removes the frequency component indistinguishable from
	an unknown complex amplitude. For conditionally locally unbiased estimation, $J_x>0$, and finite
	$\E[J_x^{-1}]$, the averaged conditional CRB is
	\begin{equation}
		\E_{\bm u}[\CRB_k]
		=\frac{\sigma^2}{8\pi^2|\alpha_k|^2}\E_{\bm u}[J_x^{-1}].
		\label{eq:expected_crb}
	\end{equation}
	The inverse must be taken before averaging, and thus replacing $\E[J_x^{-1}]$ by
	$1/\E[J_x]$ is generally incorrect. This bound is a local, isolated-target
	reference and does not include undesired-scatterer bias or incorrect peaks.

	\subsection{Local Errors and Incorrect Acquisition}
	For analysis, choose a tolerance $\Delta_{\rm tol}>0$ and define the region
	$\mathcal B_k=[\tau_k-\Delta_{\rm tol},\tau_k+\Delta_{\rm tol}]$
	and event $\mathcal E_k=\{\widetilde\tau_k\in\mathcal B_k\}$.
	It contains at least one search-grid point and is not supplied to the estimator.
	Let $\xi_r$, $r=1,\ldots,R-1$, enumerate all grid points outside
	$\mathcal B_k$, and write
	$\pi_r=\Pr\{\widetilde\tau_k=\xi_r\}$,
	$\pi_0=\Pr\{\mathcal E_k\}$, and
	$\MSE_k^{\rm loc}
	=\E[(\widehat\tau_k-\tau_k)^2\mid\mathcal E_k]$.
	Conditioning on the acquisition outcome yields the exact MSE decomposition
	\begin{equation}
		 \E[(\widehat\tau_k-\tau_k)^2] \! = \! \pi_0 \MSE_k^{\rm loc}
		+\sum\nolimits_{r=1}^{R-1}\pi_r
		\E[(\widehat\tau_k-\tau_k)^2\mid\widetilde\tau_k=\xi_r].
		\label{eq:delay_exact}
	\end{equation}
	Using the coarse-grid displacement for each incorrect outcome then yields
	\begin{equation}
		\E[(\widehat\tau_k-\tau_k)^2] \simeq\pi_0 \MSE_k^{\rm loc}
		+\sum\nolimits_{r=1}^{R-1}\pi_r(\xi_r-\tau_k)^2.
		\label{eq:sl_global_mse_simplified}
	\end{equation}
	The total incorrect-acquisition probability is
	\begin{equation}
		P_{{\rm out},k}=1-\pi_0=\sum\nolimits_{r=1}^{R-1}\pi_r.
		\label{eq:outlier_probability}
	\end{equation}
	Minimizing $P_{{\rm out},k}$ alone need not minimize MSE: a rare distant
	peak can contribute more error than a more likely nearby one. The
	displacement weights in \eqref{eq:sl_global_mse_simplified} retain this
	distinction and connect the acquisition problem to waveform design.
	Indeed, refinement changes the squared displacement of candidate $r$ by at
	most $2|\xi_r-\tau_k|\Delta_{\rm ref}+\Delta_{\rm ref}^2$.
	Thus, \eqref{eq:sl_global_mse_simplified} applies when refinement is small
	relative to the relevant incorrect displacements. Otherwise, it retains the
	exact conditional squared error for each selected grid point.
	
	For a correctly acquired peak with $\alpha_k\ne0$ and
	$\tau_k\in \mathcal T_k$, put
	$\widehat\delta=\widehat\tau_k-\tau_k$ and
	$L(\delta)=|s(\tau_k+\delta)|^2$.
	Let $L_0(\delta)$ denote the profile energy due to the desired
	target alone. Taylor expansion gives
	\begin{equation}
		\widehat\delta\simeq-\frac{L'(0)}{L_0''(0)},
		\label{eq:sl_peak_perturbation}
	\end{equation}
The approximation requires a small delay error, a refined peak away from the search boundaries, and $L''(0)\simeq L_0''(0)<0$. The closed-form MSE expressions assume $\Pr(\mathcal E_k)\simeq1$; otherwise, evaluate $\MSE_k^{\rm loc}$ numerically conditional on $\mathcal E_k$. The exact conditional MSE always satisfies $\MSE_k^{\rm loc}\le(\Delta_{\rm tol}+\Delta_{\rm ref})^2$.

	\subsection{MF: Delay-Resolved Scatterers}
	Define $P=\sum_n p_n$, the frequency centroid
	$\bar f_p=P^{-1}\sum_n p_nf_n$, and
	\begin{equation}
		\widetilde f_{p,n}=f_n-\bar f_p,\qquad
		J_p=\sum_n p_n\widetilde f_{p,n}^{\,2}.
		\label{eq:spread}
	\end{equation}
	The local MSE averages over payloads and noise at fixed scatterer
	phases. The expectation $\E_{\bm\phi}$ additionally averages over
	independent uniform undesired-scatterer phases, with magnitudes and
	delays fixed and the target phase as reference.
	
	\begin{theorem}
		\label{thm:sl_local_mf_delay_resolved}
		Under the preceding local-perturbation and reliable-acquisition
		conditions, assume negligible weighted ambiguity responses at
		the undesired delays, concentration of $J_x$ about $J_p$ with
		controlled inverse moments, and negligible coherent sidelobe
		contributions relative to the retained terms. Then, it follows that
		\begin{equation}
			\E_{\bm\phi}[\MSE_{\MF,k}^{\rm loc}]
			\simeq
			\frac{\sigma^2J_p+
				\bigl(\sum_{\ell\ne k}|\alpha_\ell|^2\bigr)
				\sum_n\kappa_np_n^2\widetilde f_{p,n}^{\,2}}
			{8\pi^2|\alpha_k|^2J_p^2}.
			\label{eq:sl_local_mf_phase_average}
		\end{equation}
	\end{theorem}
	
	\begin{proof}
		For $\ell\ne k$, define
		$\psi_{k\ell,n}=e^{j2\pi f_n(\tau_k-\tau_\ell)}$,
		$\bar\psi_{k\ell}=P^{-1}\sum_n p_n\psi_{k\ell,n}$, and
		$D_{k\ell}=\sum_n p_n\widetilde f_{p,n}\psi_{k\ell,n}$.
		Using $P_x$, $\bar f_x$, and $J_x$ from
		\eqref{eq:conditional_information}, let
		$D_{k\ell}^x=\sum_n p_n|u_n|^2(f_n-\bar f_x)\psi_{k\ell,n}$
		and
		$Z_x=\sum_n\sqrt{p_n}u_n^*z_ne^{j2\pi f_n\tau_k}
		(f_n-\bar f_x)$.
		Centering the profile at $\bar f_x$ and applying
		\eqref{eq:sl_peak_perturbation} to first order give
		\begin{equation}
			\widehat\delta_{\MF}\simeq-
			\frac{\Imm\!\left\{\alpha_k^*
				\left(\sum_{\ell\ne k}\alpha_\ell D_{k\ell}^x+Z_x\right)\right\}}
			{2\pi|\alpha_k|^2J_x}.
			\label{eq:mf_instant_error}
		\end{equation}
		Since $Z_x\mid\bm u\sim\CN(0,\sigma^2J_x)$, averaging over
		noise and the independent uniform phases yields
		\begin{align}
			\E_{\bm\phi}[\MSE_{\MF,k}^{\rm loc}]
			\! \simeq \!
			\frac{1}{8\pi^2|\alpha_k|^2}
			\E_{\bm u}\Bigg[ \!
			\frac{\sigma^2}{J_x}  \! + \! 
			\sum_{\ell\ne k}|\alpha_\ell|^2
			\frac{|D_{k\ell}^x|^2}{J_x^2}
			\! \Bigg].
			\label{eq:mf_finite_n}
		\end{align}
		
		Under the delay-resolution approximation, $|\bar{\psi}_{k\ell}|\simeq 0$.
		Linearizing the payload-dependent centering and using the
		independent symbol-energy fluctuations with variances
		$\kappa_n$ give
		\[
		\E_{\bm u}[|D_{k\ell}^x|^2]
		\simeq |D_{k\ell}|^2+
		\sum\nolimits_n\kappa_np_n^2\widetilde f_{p,n}^{\,2}.
		\]
		The concentration assumption, including controlled inverse-spread
		moments, permits replacing $J_x$ by $J_p$ in
		\eqref{eq:mf_finite_n}.
		Negligible coherent sidelobe effects mean that
		$\sum_{\ell\ne k}|\alpha_\ell|^2|D_{k\ell}|^2$
		is small relative to the retained noise and payload terms.
		Dropping this contribution gives
		\eqref{eq:sl_local_mf_phase_average}.
	\end{proof}
	
	Concentration requires small relative power fluctuations, no dominant
	tone, and controlled inverse spread. A finite alphabet with symbol energies
	bounded away from zero controls $J_p/J_x$. This frozen-denominator result
	is not a complete second-order finite-$N$ expansion.
	
	\begin{corollary}
		\label{cor:sl_local_mf_uniform}
		Under Theorem~\ref{thm:sl_local_mf_delay_resolved}, if $p_n=p$ and $\kappa_n=\kappa$ for all active tones, then
		\begin{equation}
			\E_{\bm\phi}[\MSE_{\MF,k}^{\rm loc}]\simeq
			\frac{\sigma^2/p+\kappa\sum_{\ell\ne k}|\alpha_\ell|^2}
			{8\pi^2|\alpha_k|^2\sum_n\widetilde f_n^{\,2}},
			\label{eq:sl_local_mf_uniform}
		\end{equation}
		where $\widetilde f_n=f_n-N^{-1}\sum_m f_m$.
	\end{corollary}
	\begin{proof}
		Substituting $J_p=p\sum_n\widetilde f_n^{\,2}$ and $\sum_n\kappa_np_n^2\widetilde f_{p,n}^{\,2}=\kappa p^2\sum_n\widetilde f_n^{\,2}$ into \eqref{eq:sl_local_mf_phase_average} yields \eqref{eq:sl_local_mf_uniform}.
	\end{proof}
	The payload-induced MF term vanishes for constant-modulus symbols.
	Delay separation alone does not guarantee a negligible coherent term.
	For fixed powers and scatterer amplitudes, the noise-independent $\kappa$
	term can leave the high-SNR residual illustrated in Fig.~\ref{fig:mse_regimes}.

	\subsection{RF Local Error}
	Let $\widetilde f_{0,n}=f_n-N^{-1}\sum_m f_m$,
	$J_0=\sum_n\widetilde f_{0,n}^{\,2}$, and
	$D_{k\ell}^{\RF}=\sum_n\widetilde f_{0,n}
	e^{j2\pi f_n(\tau_k-\tau_\ell)}$.
	The equalized noise $v_n=z_n/(\sqrt{p_n}u_n)$ is proper and has
	average variance $\sigma^2\eta_n/p_n$.

\begin{theorem}
	\label{thm:sl_local_rf}
	Under the preceding local-perturbation and reliable-acquisition
	conditions, with $\eta_n<\infty$ and negligible coherent sidelobe
	contributions relative to the noise term, the RF local MSE satisfies
	\begin{equation}
		\E_{\bm\phi}[\MSE_{\RF,k}^{\rm loc}]
		\simeq
		\frac{\sigma^2\sum_n\widetilde f_{0,n}^{\,2}\eta_n/p_n}
		{8\pi^2|\alpha_k|^2J_0^2}.
		\label{eq:sl_local_rf}
	\end{equation}
\end{theorem}
	\begin{proof}
		The desired RF component has curvature
		$L_0''(0)=-8\pi^2|\alpha_k|^2NJ_0$.
		Let $\widetilde v_{k,n}=v_ne^{j2\pi f_n\tau_k}$; this deterministic
		phase rotation preserves the noise moments. The first-order peak perturbation is
		\begin{equation}
			\widehat\delta_{\RF}\simeq-
			\frac{\Imm\{\alpha_k^*(\sum_{\ell\ne k}\alpha_\ell D_{k\ell}^{\RF}
				+\sum_n\widetilde f_{0,n}\widetilde v_{k,n})\}}
			{2\pi|\alpha_k|^2J_0}.
			\label{eq:rf_instant_error}
		\end{equation}
		Properness and independent subcarrier noises give
		$\E[|\sum_n\widetilde f_{0,n}\widetilde v_{k,n}|^2]
		=\sigma^2\sum_n\widetilde f_{0,n}^{\,2}\eta_n/p_n$.
		The noise imaginary-part variance is $|\alpha_k|^2/2$ times this
		value. Averaging over the independent scatterer phases and neglecting
		the coherent sidelobe contribution under the stated approximation
		yields \eqref{eq:sl_local_rf}.
	\end{proof}
	RF has no $\kappa_n$-dependent signal weighting. With constant
	modulus and equal powers, $s^{\MF}(\tau)=p\,s^{\RF}(\tau)$ for every
	observation, and both receivers return the same delay estimate.

	\subsection{Correlated Regional Probabilities}
	For fixed scene parameters, collect $S_a=s(\xi_a)$ over the full grid $\mathcal G\cap\mathcal T_k$ into $\bm S=[S_a]_a$, retaining $r=1,\ldots,R-1$ for the incorrect candidates. For receiver $q\in\{\MF,\RF\}$, $\bm S\mid\bm u\sim\CN(\bm m^{(q)},\bm C^{(q)})$, where
	\begin{align}
		m_a^{(q)}&=\sum\nolimits_n w_n^{(q)}  \sum\nolimits_{\ell=1}^K\alpha_{\ell}
		e^{-j2\pi f_n\tau_{\ell}} e^{j2\pi f_n\xi_a},\notag\\
		C_{ab}^{(q)}&=\sigma^2\sum\nolimits_n h_n^{(q)}
		e^{j2\pi f_n(\xi_a-\xi_b)},
		\label{eq:regional_stats}
	\end{align}
	 and the two receivers have weights
	\begin{equation}
		\begin{aligned}
			(w_n^{\MF},h_n^{\MF})&=(p_n|u_n|^2,p_n|u_n|^2),\\
			(w_n^{\RF},h_n^{\RF})&=(1,(p_n|u_n|^2)^{-1}).
		\end{aligned}
		\label{eq:regional_weights}
	\end{equation}
	The covariance follows by writing the profile noise as
	$\bm S-\bm m^{(q)}=\bm H_q\bm z$, with
	$[\bm H_{\MF}]_{an}=x_n^*e^{j2\pi f_n\xi_a}$ and
	$[\bm H_{\RF}]_{an}=x_n^{-1}e^{j2\pi f_n\xi_a}$.
	Hence $\bm C^{(q)}=\sigma^2\bm H_q\bm H_q^H$.
	Because the same subcarrier noise samples contribute to all delay samples, treating the grid samples as independent would discard their correlation. Sampling in subcarrier space
	also handles an oversampled grid whose covariance has rank at most $N$.
	Let $\mathcal I_r$ be the event that the coarse rule selects incorrect
	candidate $r$, including its prescribed tie breaking. For a unique maximum,
	$\mathcal I_r=\{\bm S:|S_r|^2>\max_{a\ne r}|S_a|^2\}$.
	Its selection probability is
	\begin{equation}
		\pi_r=\E_{\bm u}[\Pr\{\mathcal I_r\mid\bm u\}],\qquad
		\pi_0=1-\sum\nolimits_{r=1}^{R-1}\pi_r.
		\label{eq:regional_probability}
	\end{equation}
	Correlated Gaussian sampling evaluates the inner probability, and payload
	sampling evaluates the outer expectation. All correct-region grid points
	participate in the competition. One representative correct sample generally
	does not reproduce the implemented coarse search. The conditional Gaussian
	model is exact, whereas probability integration and
	\eqref{eq:sl_global_mse_simplified} make the final MSE semi-analytical.
	For phase-ensemble performance, average the complete fixed-scene MSE,
	including its probabilities, over $\bm\phi$.
	
	For pairwise comparison, choose $\xi_0\in\mathcal G\cap\mathcal T_k\cap\mathcal B_k$ and set $U_a=|S_a|^2$. Here, $0$ labels one reference sample, while $\pi_0$ refers to the entire correct region. Suppressing the receiver superscript, the conditional mean and variance of $\Delta_r=U_0-U_r$ are
	$
		\mu_r=|m_0|^2-|m_r|^2$ and $
		V_r=C_{00}^2+C_{rr}^2+2C_{00}|m_0|^2+2C_{rr}|m_r|^2
		-2|C_{0r}|^2-4\Ree\{m_0^*C_{0r}m_r\}.
	$
	Here $C_{00}=C_{rr}$, and the last two terms follow from
	$\Cov(U_0,U_r\mid\bm u)=|C_{0r}|^2+2\Ree\{m_0^*C_{0r}m_r\}$.
	For $V_r>0$, a moment-matched real Gaussian gives
	\begin{equation}
		\Pr\{U_r>U_0\}\approx\E_{\bm u}[Q(\mu_r/\sqrt{V_r})],
	\end{equation}
	where $Q$ is the standard Gaussian tail. This exposes how mean separation
	and correlation affect threshold behavior. The exact pairwise probability satisfies $\pi_r\leq\Pr\{U_r\geq U_0\}$, but its Gaussian approximation need not preserve this bound. Since pairwise events can overlap, evaluating \eqref{eq:sl_global_mse_simplified} requires the full-grid selection probabilities in \eqref{eq:regional_probability}.
	
	\section{Cooperative Position MSE}
	Building on Section~\ref{SingleLink}, we approximate cooperative
	position MSE by combining local fusion errors with nonlinear
	fusion costs for incorrect acquisitions, accounting for network
	geometry and cross-link dependence.
	
	\subsection{All-Correct Local Fusion}
	Let $E$ denote the number of sensing links. The local geometry model follows distributed-radar analysis \cite{Meng2024CooperativeTWC, Sadeghi2021Geometry}. The Jacobian
	$\bm A\in\mathbb R^{E\times D}$ has row
	$\bm a_e^T=(\bm\nu_{i,k}+\bm\nu_{j,k})^T$, where
	$\bm\nu_{i,k}=(\bm r_k-\bm b_i)/\|\bm r_k-\bm b_i\|$.
	Let $\mathcal E_k^{\rm all}=\bigcap_{e\in\mathcal E}\mathcal E_{e,k}$
	and $\bm\varepsilon_\tau=\widehat{\bm\tau}_k-\bm\tau_k$.
	The correctly acquired joint delay-error moments are respectively denoted by
	$
		\bm b_\tau=\E[\bm\varepsilon_\tau\mid\mathcal E_k^{\rm all}]$ and $
		\bm\Sigma_\tau=\Cov(\bm\varepsilon_\tau\mid\mathcal E_k^{\rm all}).
	$ Under reliable acquisition, these moments are approximated using the link-wise MF/RF delay-error expressions in Section \ref{SingleLink}, retaining cross-link correlations.
	If the solution is interior to $\Omega$, the local linearization is accurate,
	and $\bm A^T\bm W\bm A$ is nonsingular, then
	\begin{align}
		\vspace{-1mm}
		\widehat{\bm r}_k-\bm r_k&\simeq c\bm T
		(\widehat{\bm\tau}_k-\bm\tau_k),\notag\\
		\bm T&=(\bm A^T\bm W\bm A)^{-1}\bm A^T\bm W,
		\vspace{-1mm}
		\label{eq:local_map}
	\end{align}
	and the conditional position MSE is
	\begin{equation}
		\vspace{-1mm}
		M_k^{\rm pos,loc}\simeq c^2\tr(\bm T\bm\Sigma_\tau\bm T^T)
		+c^2\|\bm T\bm b_\tau\|^2.
		\vspace{-1mm}
		\label{eq:local_position}
	\end{equation}
	Specifically, $\bm d(\bm r_k+\Delta\bm r)\simeq
	\bm d(\bm r_k)+\bm A\Delta\bm r$ turns the fusion residual into
	$c\bm\varepsilon_\tau-\bm A\Delta\bm r$.
	Setting the gradient of its weighted squared norm to zero gives
	\begin{equation}
		\bm A^T\bm W\bm A\,\Delta\widehat{\bm r}
		=c\bm A^T\bm W\bm\varepsilon_\tau.
		\label{eq:normal_equation}
	\end{equation}
	Solving this normal equation proves \eqref{eq:local_map}; the conditional
	bias--variance identity then gives \eqref{eq:local_position}.
	For $\bm b_\tau=\bm0$ and $\bm\Sigma_\tau\succ0$, choosing $\bm W=(c^2\bm\Sigma_\tau)^{-1}$ gives
	$
	M_k^{\rm pos,loc}\simeq
	\tr\!\left[
	\left(\bm A^T(c^2\bm\Sigma_\tau)^{-1}\bm A\right)^{-1}
	\right].
	$
	Shared MF payloads may create off-diagonal covariances, which must then be
	retained. For a fixed scene, use fixed-scene conditional moments, obtainable
	from \eqref{eq:mf_instant_error} and \eqref{eq:rf_instant_error} under reliable
	acquisition. Phase-averaged scalar MSEs alone do not specify the joint
	fixed-scene covariance.
	
	\begin{figure*}[t]
		\centering
		\subfloat[QPSK.]{\includegraphics[width=.32\textwidth]{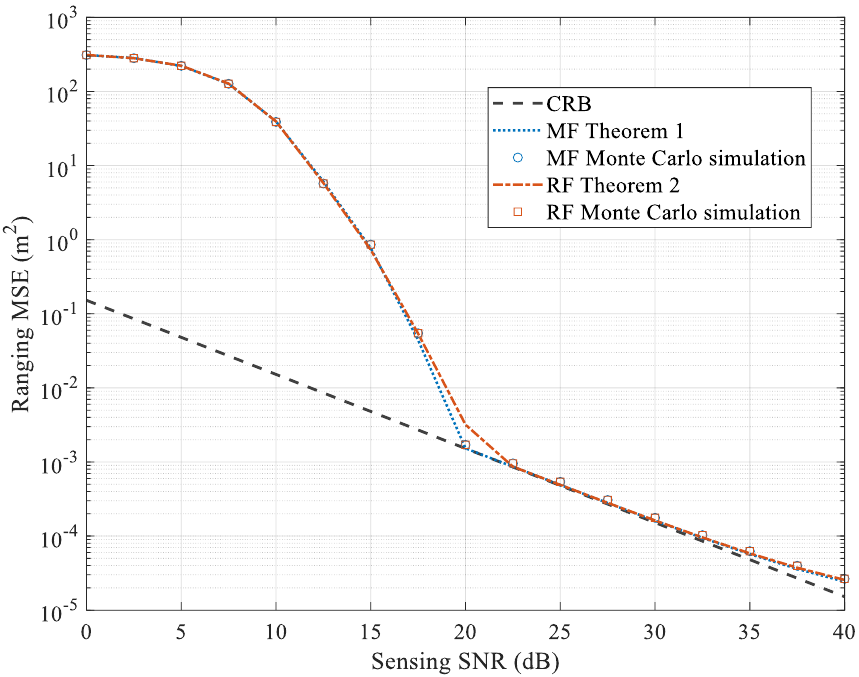}}\hfill
		\subfloat[16QAM.]{\includegraphics[width=.32\textwidth]{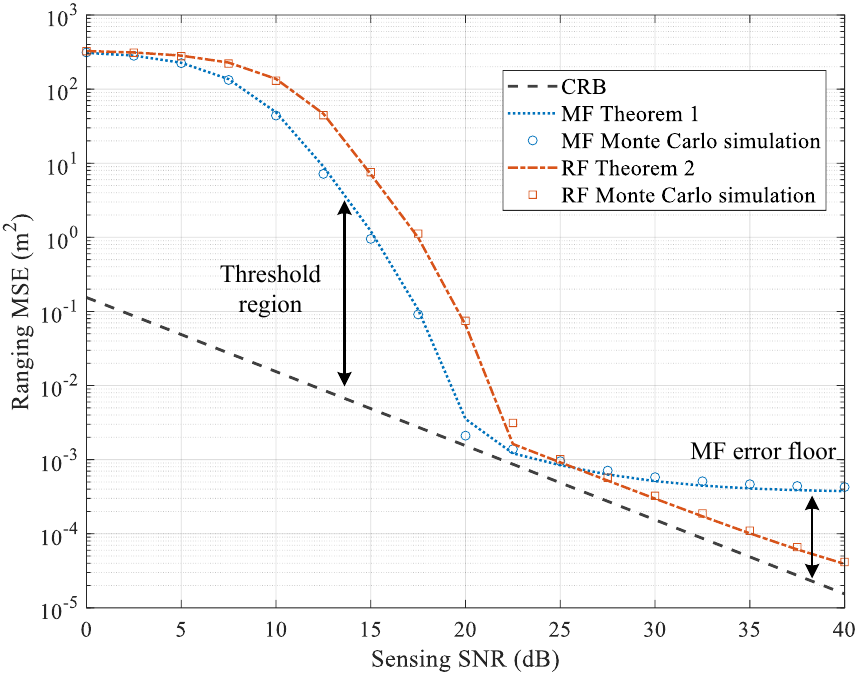}}\hfill
		\subfloat[Mixed QPSK/16QAM.]{\includegraphics[width=.32\textwidth]{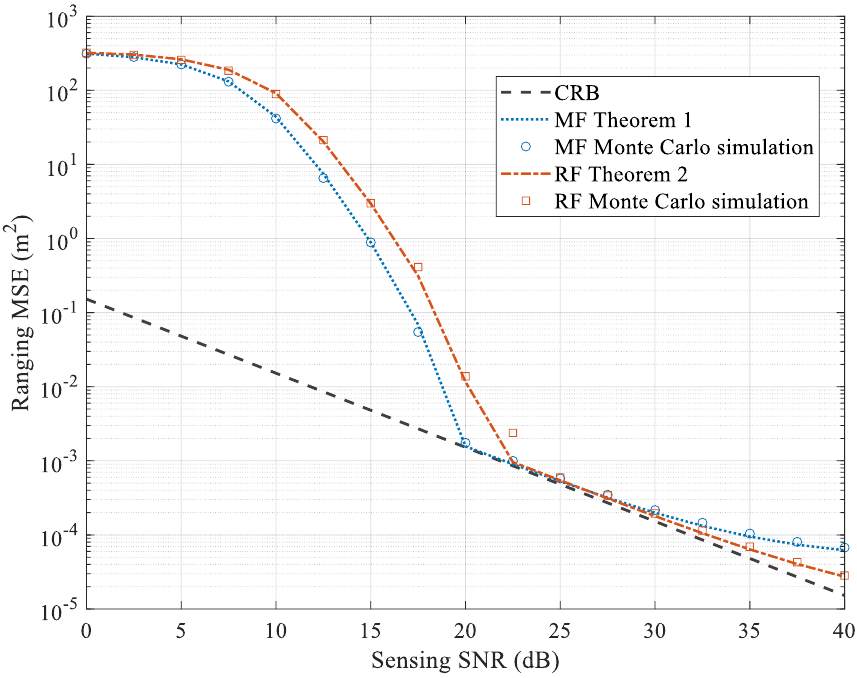}}
		\vspace{-1mm}
		\caption{Single-link propagation-length MSE versus sensing SNR for
			MF and RF with $N=64$ and $B=300\,\mathrm{MHz}$: (a) QPSK;
			(b) 16QAM; (c) QPSK on the outer half of subcarriers and 16QAM on the
			central half. }
		\label{fig:spread_mse}
	\end{figure*}
	\subsection{Incorrect-Region Displacement Costs}
	Let $s_e=0$ denote correct acquisition on link $e$, and $s_e=r>0$ selection
	of $\xi_{e,r}$. For fixed scene parameters and joint state
	$\bm s\in\mathcal S=\prod_e\{0,\ldots,R_e-1\}$,
	conditional independence across links given the payloads yields
	\begin{equation}
		\Pi_k(\bm s)=\E_{\{\bm u_i\}}\!\left[
		\prod_{e=1}^E\pi_{e,s_e}(\bm u_{i_e})\right].
		\label{eq:joint_probability}
	\end{equation}
	Here $\pi_{e,r}(\bm u_{i_e})$ is a conditional regional probability.
	The payload expectation must remain outside the product when links share a
	transmitter. Replacing it by a product of marginal probabilities requires an
	additional independence approximation, and it is exact for constant-modulus
	payloads or one sensing link per transmitting BS under the stated model.
	
	Associate each state with nominal lengths and a nonlinear-fusion position:
	\begin{align}
		\vspace{-1mm}
		[\bm d_k^0(\bm s)]_e&=\begin{cases}
			d_{e,k},&s_e=0,\\c\xi_{e,s_e},&s_e>0,
		\end{cases}\notag\\
		\bm r_k^\star(\bm s)&=\arg\min_{\bm r\in\Omega}
		\|\bm d_k^0(\bm s)-\bm d(\bm r)\|_{\bm W}^2,
		\vspace{-1mm}
		\label{eq:pseudotrue}
	\end{align}
	with cost $D_k^{\rm pos}(\bm s)=\|\bm r_k^\star(\bm s)-\bm r_k\|^2$.
	Assume the noiseless all-correct problem uniquely identifies $\bm r_k$.
	Conditioning on the state gives the exact identity
	\begin{equation}
		M_k^{\rm pos}=\sum\nolimits_{\bm s\in\mathcal S}\Pi_k(\bm s)
		\E[\|\widehat{\bm r}_k-\bm r_k\|^2\mid\bm s].
		\label{eq:exact_position}
	\end{equation}
	Let $V_{\bm s}=\E[\|\widehat{\bm r}_k-
	\bm r_k^\star(\bm s)\|^2\mid\bm s]$ denote the mean-square
	deviation from the nominal fusion position.
	Expanding the squared norm and applying Cauchy--Schwarz give
	\begin{equation}
		\left|\E[\|\widehat{\bm r}_k-\bm r_k\|^2\mid\bm s]
		-D_k^{\rm pos}(\bm s)\right|
		\leq 2\sqrt{D_k^{\rm pos}(\bm s)V_{\bm s}}+V_{\bm s}.
		\label{eq:state_cost_error}
	\end{equation}
	Thus, if $V_{\bm s}\ll D_k^{\rm pos}(\bm s)$ for each incorrect
	acquisition state $\bm s\ne\bm0$, then
	\begin{equation}
		M_k^{\rm pos}\simeq\Pi_k(\bm0)M_k^{\rm pos,loc}
		+\sum\nolimits_{\bm s\ne\bm0}\Pi_k(\bm s)D_k^{\rm pos}(\bm s).
		\label{eq:global_position}
	\end{equation}
	For incorrect states violating this condition, retain their
	numerically evaluated conditional MSEs in \eqref{eq:exact_position}.
	The displacement costs capture nonlinear fusion effects and can
	be precomputed for fixed geometry, grid, and weights.
	A single-outlier truncation is appropriate only when multiple-link
	failures contribute negligibly to MSE; otherwise, use full-state
	evaluation or joint state sampling.

	\begin{figure*}[t]
		\centering
		\vspace{-1mm}
		\subfloat[Multistatic geometry.]{\includegraphics[width=.29\textwidth,height=47mm,keepaspectratio]{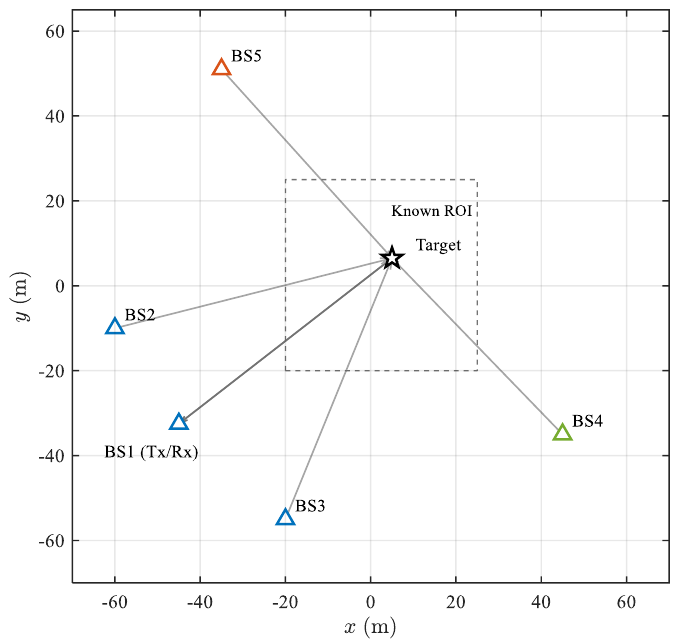}}\hfill
		\subfloat[MF.]{\includegraphics[width=.32\textwidth,height=47mm,keepaspectratio]{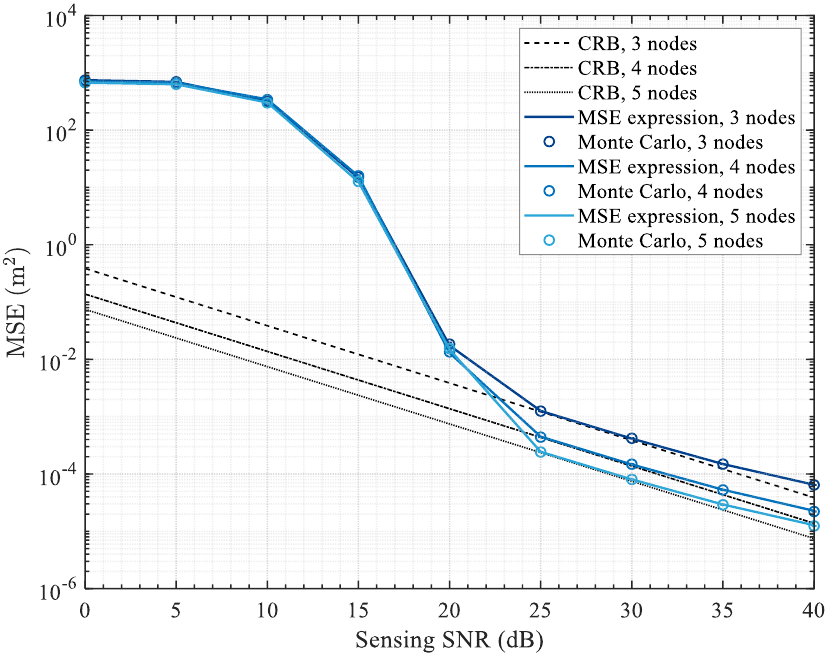}}\hfill
		\subfloat[RF.]{\includegraphics[width=.32\textwidth,height=47mm,keepaspectratio]{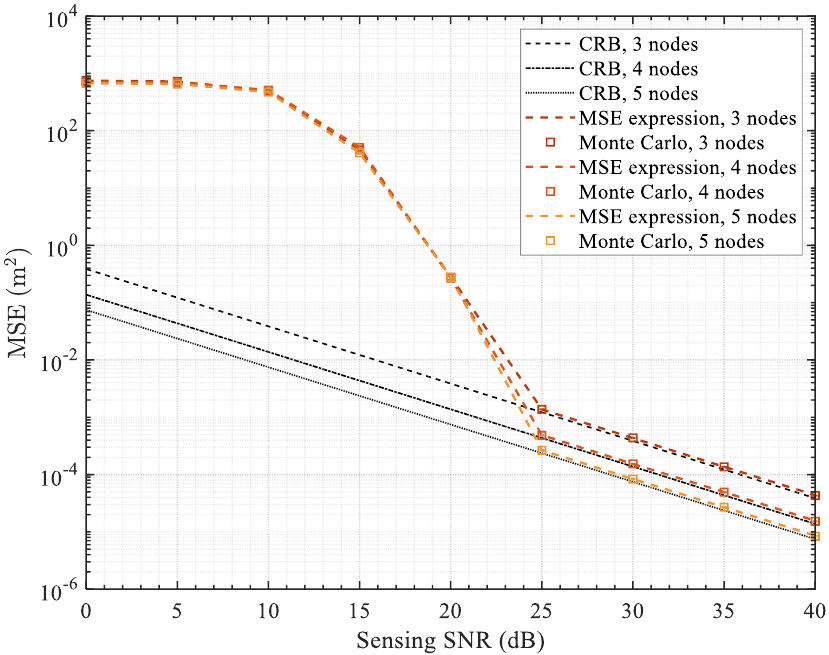}}
		\vspace{-1.5mm}
		\caption{Cooperative localization with three, four, and five BSs:
			(a) multistatic geometry; (b) MF position MSE; (c) RF position MSE.}
		\label{fig:localization_mse}
	\end{figure*}

	\section{Power Control With Fixed Constellations}
	
	\subsection{Feasible Powers}
	For fixed constellations, $\kappa_n$ and $\eta_n$ are constants
	\cite{Meng2026Constellation}. 
	The lower bounds satisfy
	$\underline p_n\geq
	\Gamma_n\sigma_{\rm com}^2/|h_n^{\rm com}|^2$,
	where $\Gamma_n$, $h_n^{\rm com}\ne0$, and $\sigma_{\rm com}^2$
	denote the communication SNR threshold, channel, and noise variance.
	We optimize local delay-MSE surrogates and evaluate position MSE
	using \eqref{eq:global_position}.
	
	\subsection{RF: Closed-Form Power Allocation}
	Based on the analysis in Section \ref{SingleLink},
	Theorem~\ref{thm:sl_local_rf} gives
	\begin{equation}
		\min_{\bm p\in\mathcal P}\ \sum_n\frac{a_n}{p_n},
		\label{eq:rf_power_problem}
	\end{equation}
	where $ a_n=\eta_n\widetilde f_{0,n}^{\,2}$, with $p_n \ge 0$. The KKT solution is
	\begin{equation}
		p_n^\star=
		\left[\sqrt{a_n/\lambda}\right]_{\underline p_n}^{\overline p_n},
		\label{eq:rf_power_closed}
	\end{equation}
	where $[x]_l^u=\min\{u,\max\{l,x\}\},$ and $\lambda>0$ is chosen by bisection to satisfy
	$
	\sum_n p_n^\star=
	\min\left\{
	P_{\rm tot},
	\sum_{n:a_n>0}\overline p_n+
	\sum_{n:a_n=0}\underline p_n
	\right\}.
	$
	On unclipped tones,
	$p_n^\star\propto|\widetilde f_{0,n}|\sqrt{\eta_n}$.
	For fixed fusion weights, this allocation also minimizes each BS's
	RF noise contribution to local position MSE.
	
	\subsection{MF: A Convex Fixed-Centroid Subproblem}
	Let $S_{-k}=\sum_{\ell\ne k}|\alpha_\ell|^2$.
	Choose a feasible centroid $f_c$ and impose
	$\sum_n d_np_n=0$, where $d_n=f_n-f_c$ and $b_n=d_n^2$.
	Then $J=\sum_n b_np_n=J_p$, and
	Theorem~\ref{thm:sl_local_mf_delay_resolved} gives	
	\begin{equation}
		F_{\MF}(\bm p)=
		\frac{\sigma^2}{J}
		+S_{-k}\frac{\sum_n\kappa_nb_np_n^2}{J^2}.
		\label{eq:mf_power_objective}
	\end{equation}
	With $t=1/J$ and $z_n=tp_n$, the equivalent problem is
	\begin{equation}
		\begin{aligned}
			\min_{\bm z,t\geq0}\quad&
			\sigma^2t+S_{-k}\sum_n\kappa_nb_nz_n^2\\
			\mathrm{s.t.}\quad&
			\sum_n b_nz_n=1,\quad \sum_n d_nz_n=0,\\
			&\underline p_nt\leq z_n\leq\overline p_nt,\\
			&\sum_n z_n\leq P_{\rm tot}t.
		\end{aligned}
		\label{eq:mf_power_qp}
	\end{equation}
	This QP is convex, with $t>0$ and powers recovered as $p_n=z_n/t$.
	 
	Overall, RF favors tones with larger frequency offsets and inverse-energy moments. MF balances frequency spread against a quadratic interference penalty weighted by symbol-energy variance. The local-model assumptions must be checked at the resulting allocation.

	\section{Numerical Results}
\vspace{-1.5mm}
	We consider $N=64$ subcarriers and bandwidth $B=300\,\mathrm{MHz}$. We compare QPSK, uniform 16QAM, and a mixed allocation with QPSK on the outer half of the subcarriers and 16QAM on the central half to illustrate the effects of payload statistics and constellation placement on MF/RF sensing accuracy.
	In Fig.~\ref{fig:spread_mse}, incorrect-peak displacements in
	\eqref{eq:sl_global_mse_simplified} explain the threshold gap from
	the local isolated-target CRB. The curves labeled “MF Theorem 1” and “RF Theorem 2” represent the regional MSE approximation obtained from (10) and (22), with the local-error terms evaluated as described in Section III-E. With constant-modulus symbols and equal subcarrier powers $p_n=p$, $s^{\MF}(\tau)=p\,s^{\RF}(\tau)$, so both receivers yield the same delay estimate.
	Under reliable acquisition, \eqref{eq:mf_finite_n} retains any
	non-negligible coherent ambiguity residual in the MF local reference. The mixed allocation places constant-modulus QPSK at the band edges, where larger frequency offsets provide greater delay sensitivity. On the centered, equally spaced grid, the central half contributes only about one eighth of the sum of squared frequency offsets. Retaining 16QAM there confines symbol-energy fluctuations to less influential subcarriers, reducing the MF payload--scatterer error floor and RF noise enhancement relative to uniform 16QAM.
	For 16QAM, MF enters the transition earlier in this example, while RF
	achieves lower high-SNR MSE. Under the conditions of Theorem~\ref{thm:sl_local_mf_delay_resolved}, \eqref{eq:sl_local_mf_uniform} gives a high-SNR MF propagation-length MSE floor of $c^2\kappa \sum_{\ell\ne k}|\alpha_\ell|^2 /(8\pi^2|\alpha_k|^2J_0)$.
	RF removes this signal weighting but amplifies noise for low-energy symbols. In this 16QAM example, MF therefore yields lower MSE in the transition region at lower SNRs, whereas RF achieves lower MSE at high SNR.

	Fig.~\ref{fig:localization_mse} shows lower CRBs and high-SNR position MSEs as the network expands from three to five BSs, with each link using an equal mixture of QPSK and 16QAM. In the illustrated configurations, additional BSs provide complementary observation directions and more delay measurements, strengthening the geometric constraints on the target position. The achieved MSE also depends on the joint delay-error statistics. Near the acquisition threshold, incorrect link delays can still produce large position errors through nonlinear fusion, so cooperative gains depend on both local accuracy and acquisition reliability.
	
	\section{Conclusion}
	\vspace{-0.5mm}
	This paper developed a receiver-specific regional MSE framework combining local delay errors, correlated peak selection, and nonlinear cooperative fusion costs. The analysis characterizes MF payload--scatterer coupling, RF inverse-energy noise enhancement, and coherent ambiguity residuals. For fixed constellations, the local criteria yield RF power allocations under subcarrier bounds and a convex MF subproblem with a fixed frequency centroid. Numerical results illustrate receiver-dependent threshold and high-SNR behavior, together with the benefits of frequency-dependent constellation allocation and cooperative localization.

	\vspace{-1.5mm}
	% Start the bibliography in the right column. If the main text already
	% ends there, continue in that column without inserting another break.
	\makeatletter
	\bibliographystyle{IEEEtran}
	\bibliography{references}

\begin{thebibliography}{10}
\providecommand{\url}[1]{#1}
\csname url@samestyle\endcsname
\providecommand{\newblock}{\relax}
\providecommand{\bibinfo}[2]{#2}
\providecommand{\BIBentrySTDinterwordspacing}{\spaceskip=0pt\relax}
\providecommand{\BIBentryALTinterwordstretchfactor}{4}
\providecommand{\BIBentryALTinterwordspacing}{\spaceskip=\fontdimen2\font plus
\BIBentryALTinterwordstretchfactor\fontdimen3\font minus
  \fontdimen4\font\relax}
\providecommand{\BIBforeignlanguage}[2]{{%
\expandafter\ifx\csname l@#1\endcsname\relax
\typeout{** WARNING: IEEEtran.bst: No hyphenation pattern has been}%
\typeout{** loaded for the language `#1'. Using the pattern for}%
\typeout{** the default language instead.}%
\else
\language=\csname l@#1\endcsname
\fi
#2}}
\providecommand{\BIBdecl}{\relax}
\BIBdecl

\bibitem{Li2023Towards}
R.~Li, Z.~Xiao, and Y.~Zeng, ``Toward seamless sensing coverage for cellular
  multi-static integrated sensing and communication,'' \emph{IEEE Trans.
  Wireless Commun.}, vol.~23, no.~6, pp. 5363--5376, 2024.

\bibitem{Meng2024CooperativeTWC}
K.~Meng, C.~Masouros, A.~P. Petropulu, and L.~Hanzo, ``Cooperative {ISAC}
  networks: Performance analysis, scaling laws, and optimization,'' \emph{IEEE
  Trans. Wireless Commun.}, vol.~24, no.~2, pp. 877--892, 2025.

\bibitem{Zhang2021OverviewSignal}
J.~A. Zhang \emph{et~al.}, ``An overview of signal processing techniques for
  joint communication and radar sensing,'' \emph{IEEE J. Sel. Topics Signal
  Process.}, vol.~15, no.~6, pp. 1295--1315, 2021.

\bibitem{godrich2010target}
H.~Godrich, A.~M. Haimovich, and R.~S. Blum, ``Target localization accuracy
  gain in {MIMO} radar-based systems,'' \emph{IEEE Trans. Inf. Theory},
  vol.~56, no.~6, pp. 2783--2803, Jun. 2010.

\bibitem{Liu2022Limits}
A.~Liu \emph{et~al.}, ``A survey on fundamental limits of integrated sensing
  and communication,'' \emph{IEEE Commun. Surveys Tuts.}, vol.~24, no.~2, pp.
  994--1034, 2022.

\bibitem{ZivZakai1969}
J.~Ziv and M.~Zakai, ``Some lower bounds on signal parameter estimation,''
  \emph{IEEE Trans. Inf. Theory}, vol.~15, no.~3, pp. 386--391, 1969.

\bibitem{Bell1997}
K.~L. Bell, Y.~Steinberg, Y.~Ephraim, and H.~L. Van~Trees, ``Extended
  {Ziv--Zakai} lower bound for vector parameter estimation,'' \emph{IEEE Trans.
  Inf. Theory}, vol.~43, no.~2, pp. 624--637, 1997.

\bibitem{Han2026MSE}
K.~Han, K.~Meng, A.~Chatzicharistou, and C.~Masouros, ``Constellation design in
  {OFDM-ISAC} over data payloads: From {MSE} analysis to experimentation,'' in
  \emph{Proc. IEEE Int. Conf. Commun. (ICC)}, 2026.

\bibitem{Meng2026Constellation}
K.~Meng, K.~Han, C.~Masouros, and F.~Liu, ``Constellation selection and power
  control for {OFDM}-based {ISAC}: From theory to prototype,'' \emph{IEEE
  Trans. Signal Process.}, 2026.

\bibitem{Mallat2014TOA}
A.~Mallat \emph{et~al.}, ``Statistics of the {MLE} and approximate upper and
  lower bounds---{Part I}: Application to {TOA} estimation,'' \emph{IEEE Trans.
  Signal Process.}, vol.~62, no.~21, pp. 5663--5676, Nov. 2014.

\bibitem{Shi2022DeviceFree}
Q.~Shi, L.~Liu, S.~Zhang, and S.~Cui, ``Device-free sensing in {OFDM} cellular
  network,'' \emph{IEEE J. Sel. Areas Commun.}, vol.~40, no.~6, pp. 1838--1853,
  Jun. 2022.

\bibitem{Sadeghi2021Geometry}
M.~Sadeghi, F.~Behnia, R.~Amiri, and A.~Farina, ``Target localization geometry
  gain in distributed {MIMO} radar,'' \emph{IEEE Trans. Signal Process.},
  vol.~69, pp. 1642--1652, 2021.

\end{thebibliography}
\end{document}